\documentstyle[11pt,conf_iap,epsf]{article}
\begin{document}
\heading{The MACHO PROJECT: MICROLENSING AND VARIABLE STARS}

\photo{ }

\author{K.H.COOK$^{1,2}$,  C. ALCOCK$^{1,2}$, D.R. ALVES$^{1,4}$,  
R.A. ALLSMAN$^{2}$, T.S. AXELROD$^{1,10}$, A. BECKER$^{5}$,
D.P.  BENNETT$^{1,2}$,K.C. FREEMAN$^{10}$, 
K. GRIEST$^{2,6}$, J.A. GUERN$^{2,6}$, M.J. LEHNER$^{2,6}$,
S.L. MARSHALL$^{1,2}$, D. MINNITI$^{1}$, 
B.A. PETERSON$^{10}$, M.R. PRATT$^{1,5}$, 
P.J. QUINN$^{7}$, A.W. RODGERS$^{10}$, C.W. STUBBS$^{5}$, 
W. SUTHERLAND$^{8}$, D.L. WELCH$^{9}$ \\
(The MACHO Collaboration)}
       {$^{1}$ Lawrence Livermore National Laboratory, Livermore, CA 94550\\
       $^{2}$ Center for Particle Astrophysics, University of California, Berkeley, CA 94720\\
       $^{3}$ Supercomputing Facility, Australian National Univ., Canberra, ACT 0200, Australia\\
       $^{4}$ Department of Physics, University of California, Davis, CA 95616\\
       $^{5}$ Department of Physics and Astronomy, University of Washington, Seattle, WA 98195\\
       $^{6}$ Department of Physics, University of California San Diego, La Jolla, CA 92093-0350\\
       $^{7}$ European Southern Observatory, D-85748 Garching bei M\"unchen, Germany\\
       $^{8}$ Department of Physics, University of Oxford, Oxford OX1 3RH, U.K.\\
       $^{9}$ Dept. of Physics and Astronomy, McMaster Univ., Hamilton, Ontario, Canada L8S 4M1\\
       $^{10}$ Mount Stromlo and Siding Springs Obs., Australian Natl. Univ., Weston, ACT 2611, Australia}

\bigskip

\begin{abstract}{\baselineskip 0.4cm 

The MACHO Project monitors millions of stars in the Large Magellanic Cloud,
the Small Magellanic Cloud and the bulge of the Milky Way searching for
the gravitational microlensing signature of baryonic dark matter.  This
Project has yielded surprising results.  An analysis of two years of data
monitoring the Large Magellanic Cloud points to {$\sim 50\%$} of the mass
of the Milky Way's halo in compact objects of {$\sim 0.5 M_{\odot}$}.
An analysis of one year of monitoring the bulge has yielded more 
microlensing than predicted without the invocation of a massive bar or
significant disk dark matter.  The huge database of light curves created
by this search is yielding information on extremely rare 
types of astrophysical variability
as well as providing temporal detail for the study of
well known variable astrophysical phenomena.  The variable star catalog
created from this database is previewed and example light curves
are presented.

}
\end{abstract}

\section{Introduction}

The MACHO Project is monitoring millions of stars every night searching for
the gravitational microlensing signature of massive compact halo objects
(Machos).  This project has the dedicated use the Great Melbourne Telescope at 
Mount Stromlo.  This endeavor was stimulated by Paczy\'nski's suggestion 
\cite{pac86} that gravitational microlensing was a possible way to detect
baryonic dark matter in the halo of the Milky Way.  
The principle of microlensing is simple; if a Macho lies near the line
of sight to a background star (the source), it will deflect light from the
source and produce two images.  For galactic scales, these images cannot
be resolved even by HST ($ \sim 0.001$ arcsec and thus microlensing),
but the two unresolved images combine to give an
apparent increase in the source brightness.
Due to the relative motions of the observer, lens and source, this
magnification is transient, so the effect appears as a
symmetrical and unique brightening in an otherwise constant star.
The duration of the event is a function of the mass of the lens, the
relative distances of the lens and source and the motion of the lens
with respect to the line of sight.  The magnification is just a function of
the distance of the lens from the line of sight.  The probability that a
source will be microlensed is termed the optical depth to microlensing, $\tau$.

The Large Magellanic Cloud (LMC) was chosen as the primary target because its
line of sight passes through much of the halo yet it is relatively nearby,
and contains millions of stars resolvable in modest seeing with a small 
telescope.  Our secondary target was chosen as the Galactic bulge for two
reasons.  Although the LMC is circumpolar from Mount Stromlo, it is at too
high an airmass to be usefully observed when the  bulge is overhead.
Calculations before we began taking data \cite{kim91} suggested that there should be an
observable microlensing signal from known stellar populations in the disk
and the bulge so observing the bulge would serve to prove whether microlensing 
could be detected with our system.  We also observe the Small
Magellanic Cloud (SMC) at a lower priority.  The SMC provides a different
line of sight through the halo, but is sufficiently farther away and
smaller than the LMC that we can only monitor a few million stars with
our system.
  
\section{Instrumentation, Observations, and Reductions}

\begin{figure}
\begin{center}
\epsfxsize=.45\columnwidth
\begin{minipage}{\epsfxsize}
\epsfbox{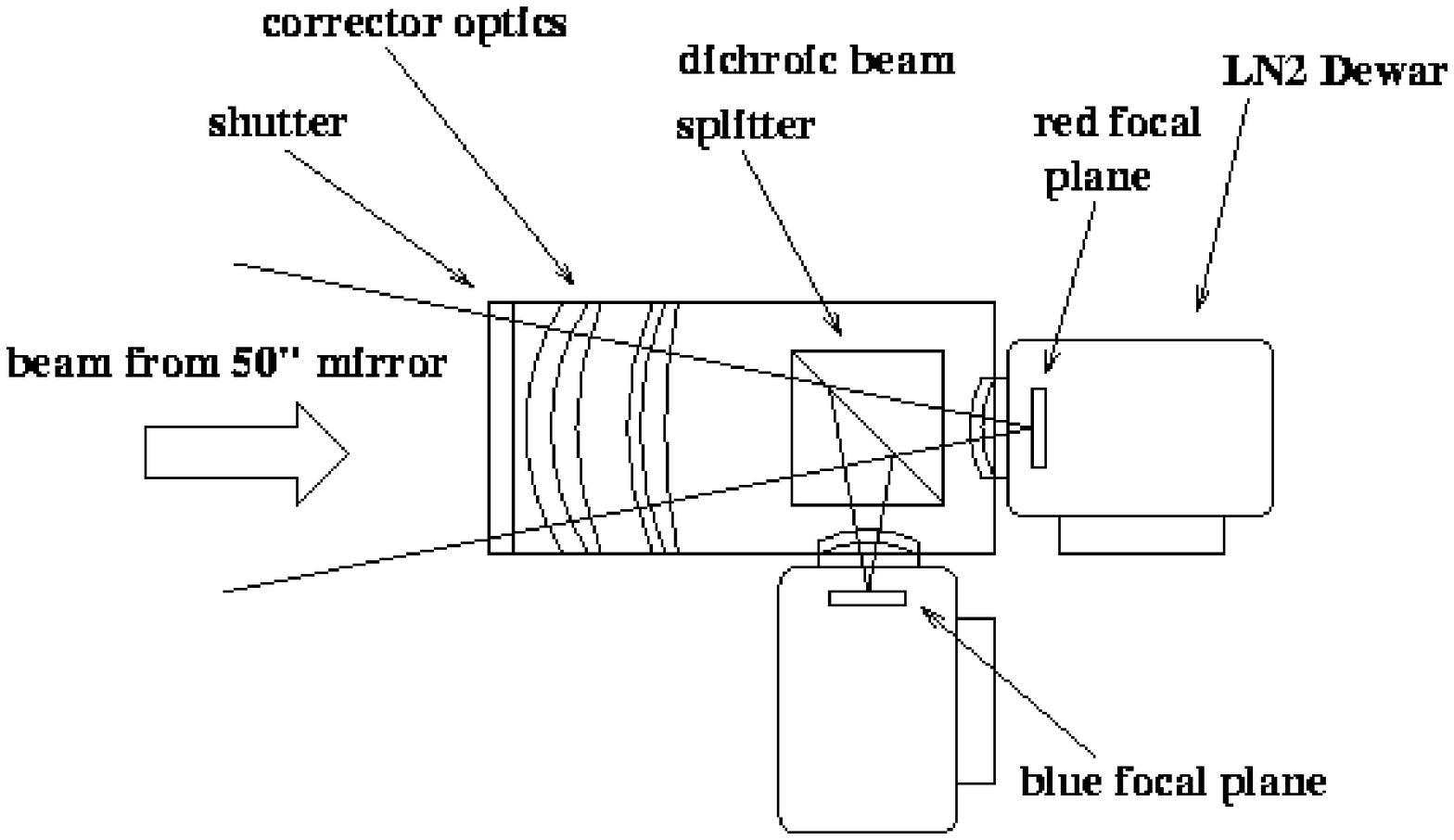}
\end{minipage}
\begin{minipage}{\epsfxsize}
\epsfbox{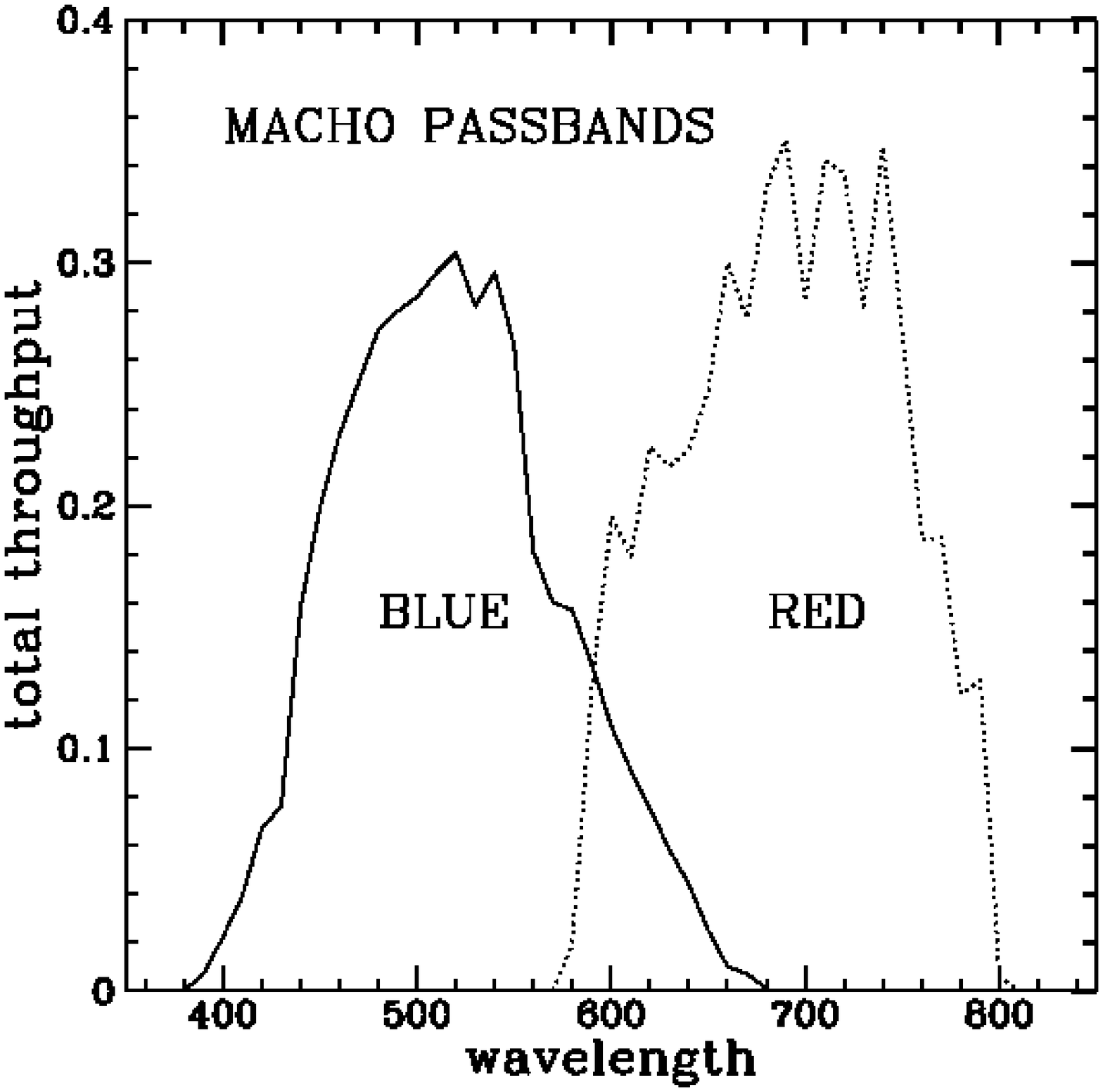}
\end{minipage}
\caption{
The left panel shows the telescope optics which re-image the {\it f}/4.5 
primary beam
to {\it f}/3.9 and produce a 1 degree corrected field.  The right panel
shows the total throughput as a function of wavelength for the red and blue
cameras of the MACHO Project.
}\label{camera}
\end{center}
\end{figure}

The refurbishment of the 1.27-m Great Melbourne Telescope has been described
by Hart {\it et al.} \cite{hart96}.  Figure~\ref{camera} shows in detail the corrector,
beam splitter and CCD cameras which allow us to monitor 0.5 square deg
simultaneously in two passbands.  These two passbands were chosen to
maximally utilize the quantum efficiency of our thick CCDs while cutting
out much of the sky brightness due to OH emission in the red and scattering
in the blue.  Figure~\ref{camera} also shows the total throughput as a function
of wavelength for our two cameras.  
Each camera contains a 2x2 mosaic of 2048x2048 CCDs.
Exposure times are 300 s toward the LMC, 150 s toward the bulge and
600 s toward the SMC.  It takes about 70 s to read out the 32 Mpixels.
Details of the camera system can be found in \cite{stubbs93}.

Images are flattened immediately upon acquisition, then simultaneously
archived to 8-mm tape and photometry begun.  Reductions and analysis are done
in the dome using two Sparc 1000s and a large disk farm (currently 
$\sim$~200 Gbyte).  MACHO field centers are defined
so that the 0.5 square degree fields tile the denser regions of the LMC, SMC 
and much of the southeast quadrant of the bulge within 10 degrees of 
the Galactic center.  These field centers (as well as much additional
information on the MACHO Project) can be found at URL 
http://wwwmacho.anu.edu.au.  Our observing strategy is to sequentially
image each of these fields, spiraling outward from dense bar fields in
the LMC and generally moving from west to east in
the bulge. As of September 1996, over 48,000 exposures have
been taken with the system, of which about $60\%$ are of the LMC.
The data covered in this presentation is derived from the
central 22 fields in the LMC observed for the first two years of the project
(September 1992 to November 1994)
and a central 24 fields in the bulge observed in the project's
first bulge season (1993).  

Deep images from the early months of the project have been reduced to
create photometry templates.  A template is divided into 64 `chunks'
in each color and
relatively bright, unblended stars in each chunk are chosen
as fiducials to define the point spread function (PSF) and 
astrometric reference frame in
future observations. A routine observation is reduced chunk by chunk
by determining the geometric transformation from the template to the current
chunk using the fiducial stars, defining a PSF using the fiducials,
and fitting at each transformed template position (more details can
be found in \cite{lmc1} and \cite{alc96}).  This scheme allows rapid
reductions, and a whole night of observations can be reduced within a few
hours of dawn.  Although the microlensing search is effected using only 
relative photometry, it is useful to derive transformations to standard
astronomical passbands.  This is complicated by two systematic effects:
1) the template observation for each field is subject to different extinction,
seeing, and possible cloud obscuration, and 2) the photometric zero point
varies from chunk to chunk (see \cite{alc96}) 
within a template observation due to
systematic effects of the photometry code. (This is due to the fact that
there is no effort made to derive an aperture correction for individual
chunk reductions.)   We have obtained roughly 30
photometric nights of CCD data using the CTIO 0.9-m and SSO 1-m to measure
the necessary 1408 transformations (64 chunks per field and 22 fields).
Although the analysis of this data is not complete, we have determined a
mean transformation to Johnson V and Kron-Cousins R which we will use in this
paper.  This transformation yields magnitudes good to about $\pm 0.15$ mag
and colors which are good to about $\pm 0.1$ mag (the zero point offsets
in the two passbands are correlated).
Light curves of time-ordered photometry are then built for
all template objects.  Light curves are not created for objects
which are not in the template. These light curves are then analyzed for
evidence of microlensing \cite{lmc1} \cite{bulge45} \cite{alc96}.
Because photometry is available within hours of an observation, a star's
current magnitude can be compared to its mean magnitude determined from
a previous analysis and if a previously constant star demonstrates a
significant deviation, our software produces an alert.  These alerts
have been used to
notify observers around the world that microlensing events are in
progress (see http://darkstar.astro.washington.edu).

\begin{center} 
{\bf Table 1.} Optical Depth Estimates
\end{center}
\begin{center}
\begin{tabular}{|l|l|l|}
\hline 
Data Set & $\tau (10^{-7})$ & 95 \% CL  \\
\hline
LMC:   8 events &  2.93  & 1.47 --- 5.28  \\
LMC:   6 events &  2.06 &  0.93 --- 4.00  \\
LMC:   no short events & $<$ 0.4  & 0.00 --- 0.4  \\
Bulge:  all stars  & 24.3  & 16.9 --- 33.3  \\
Bulge:  clump giants & 39.2 & 29.0 --- 70.9  \\
\hline
\end{tabular}
\end{center}

\begin{figure}[t]
\begin{center}
\epsfxsize=0.7\columnwidth
\begin{minipage}{\epsfxsize}
\epsfbox{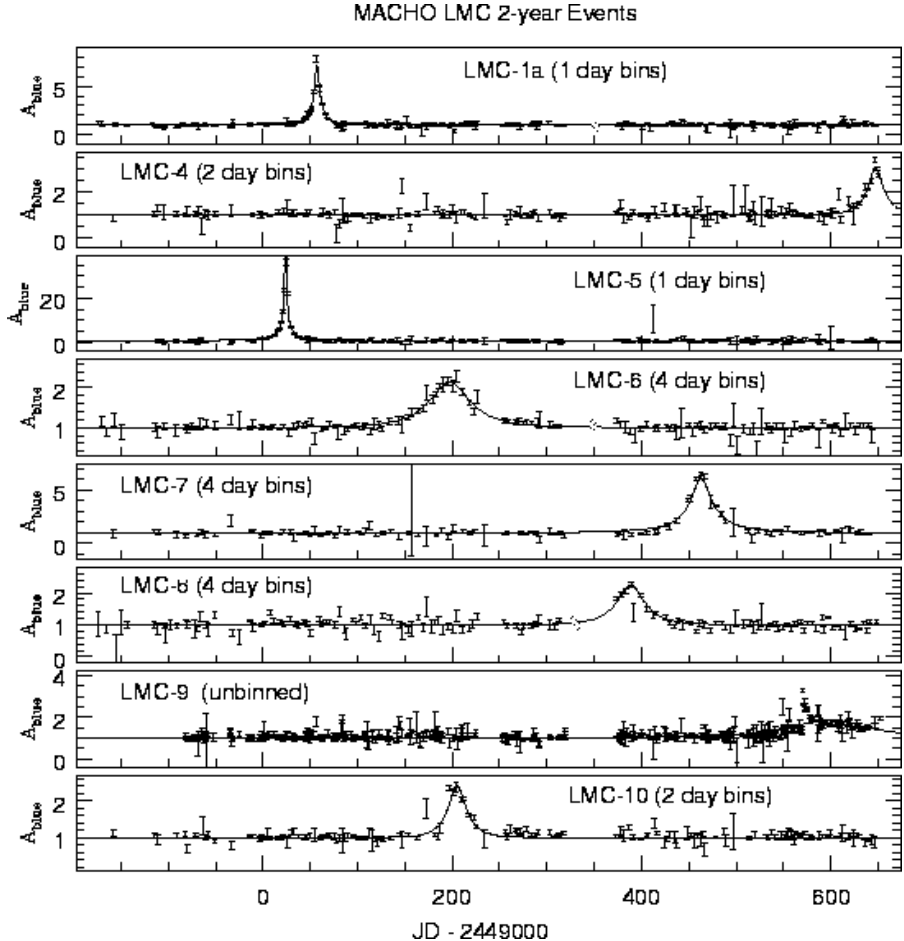}
\end{minipage}
\caption{
The 8 LMC microlensing events are plotted as time versus 
relative blue camera flux normalized to the mean 
unlensed flux (A$_{blue}$).  The solid line is the fit to simple 
microlensing and the data has been averaged in bins appropriate 
to the event duration.  LMC-9 is due to lensing by a binary as
described in \protect\cite{lmcbinary}.
}\label{lmcmicro}
\end{center}
\end{figure}

\section{Microlensing Results}

We have detected significant microlensing toward both the LMC and the
bulge.  Our microlensing discoveries are important for three reasons:
1) we have detected significant microlensing by halo dark matter looking toward
the LMC, 2) we have detected no short 1 microlensing events
looking toward the LMC and so have ruled out a significant fraction
of the halo being composed of objects with mass
$2.5 \times 10^{-6} - 8.0 \times  10^{-2} M_{\odot}$,
and 3) we have detected a higher optical depth for
microlensing toward the bulge than predictions based on simple Milky Way
models, lending support to the presence of a significant bar.

It is important to note that the microlensing interpretation for the events
we report has been strongly supported and can be considered proven.
The majority of events are well fit by the simple functional form needed
to describe point source, point lens microlensing.  There have been
spectroscopic observations through the course of microlensing toward
the bulge \cite{benetti} and the LMC \cite{lmcspec} which have shown
no spectral changes during apparent brightness changes of more 
than one magnitude 
in the span of a week.  The distribution of magnifications has been
shown to be that expected from the uniform distribution of impact parameters
which result from microlensing \cite{bulge45}.  Exceptions to simple lensing 
support the microlensing interpretation because they are predicted deviations
due to the motion of the earth \cite{parallax}, due to the finite size of
stars \cite{finite}, and due to the existence of 
binary lenses \cite{oglebinary} \cite{lmcbinary}.
While there are deviations from simple lensing due to the blending of lensed
and unlensed images, this is well understood and can be accounted for by
fitting an unmagnified fraction of the baseline flux, and thus adding
an additional fit parameter in each color \cite{lmc1} \cite{alc96}.

\begin{figure}
\begin{center}
\epsfxsize=.45\columnwidth
\begin{minipage}{\epsfxsize}
\epsfbox{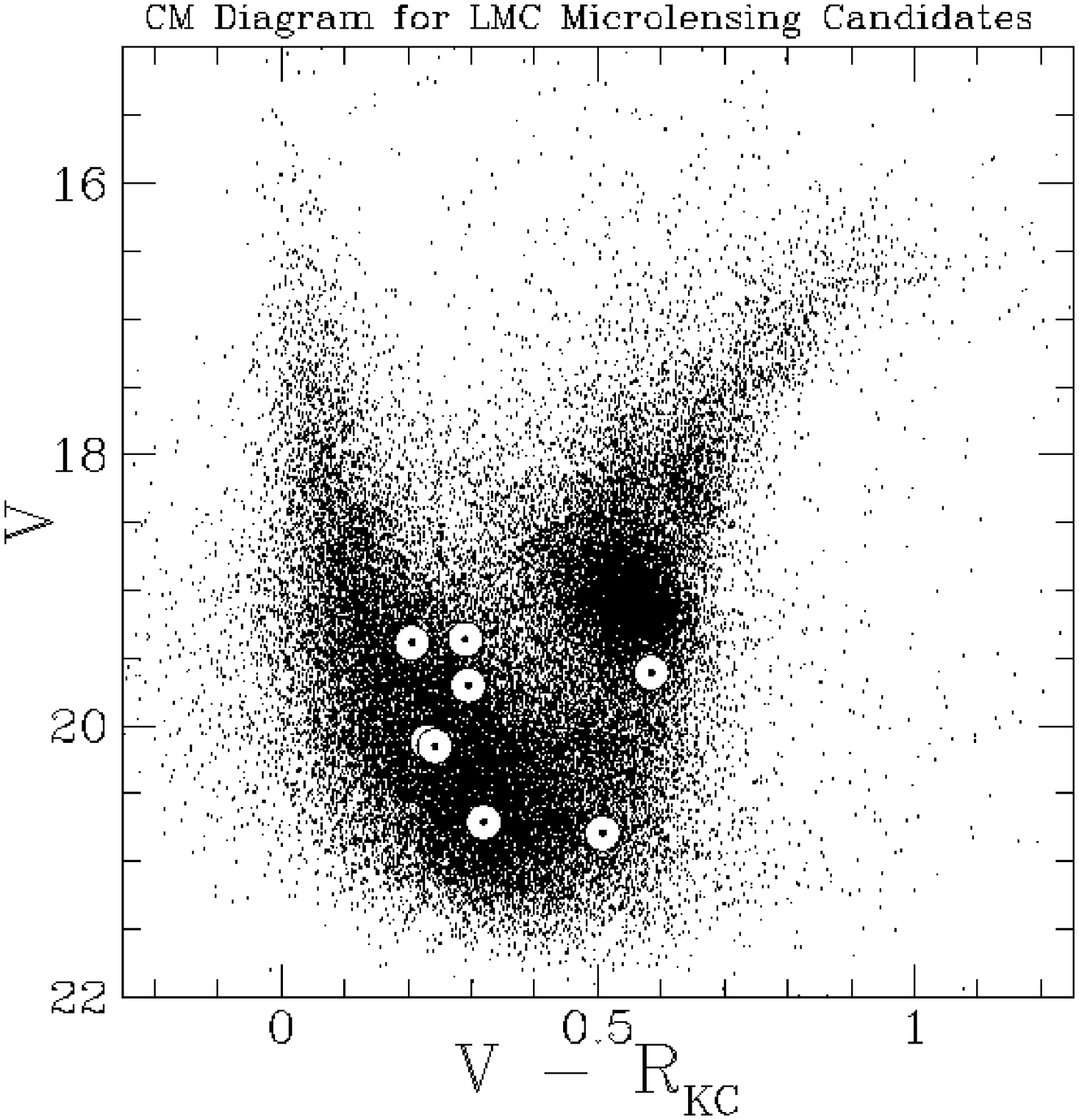}
\end{minipage}
\begin{minipage}{\epsfxsize}
\epsfbox{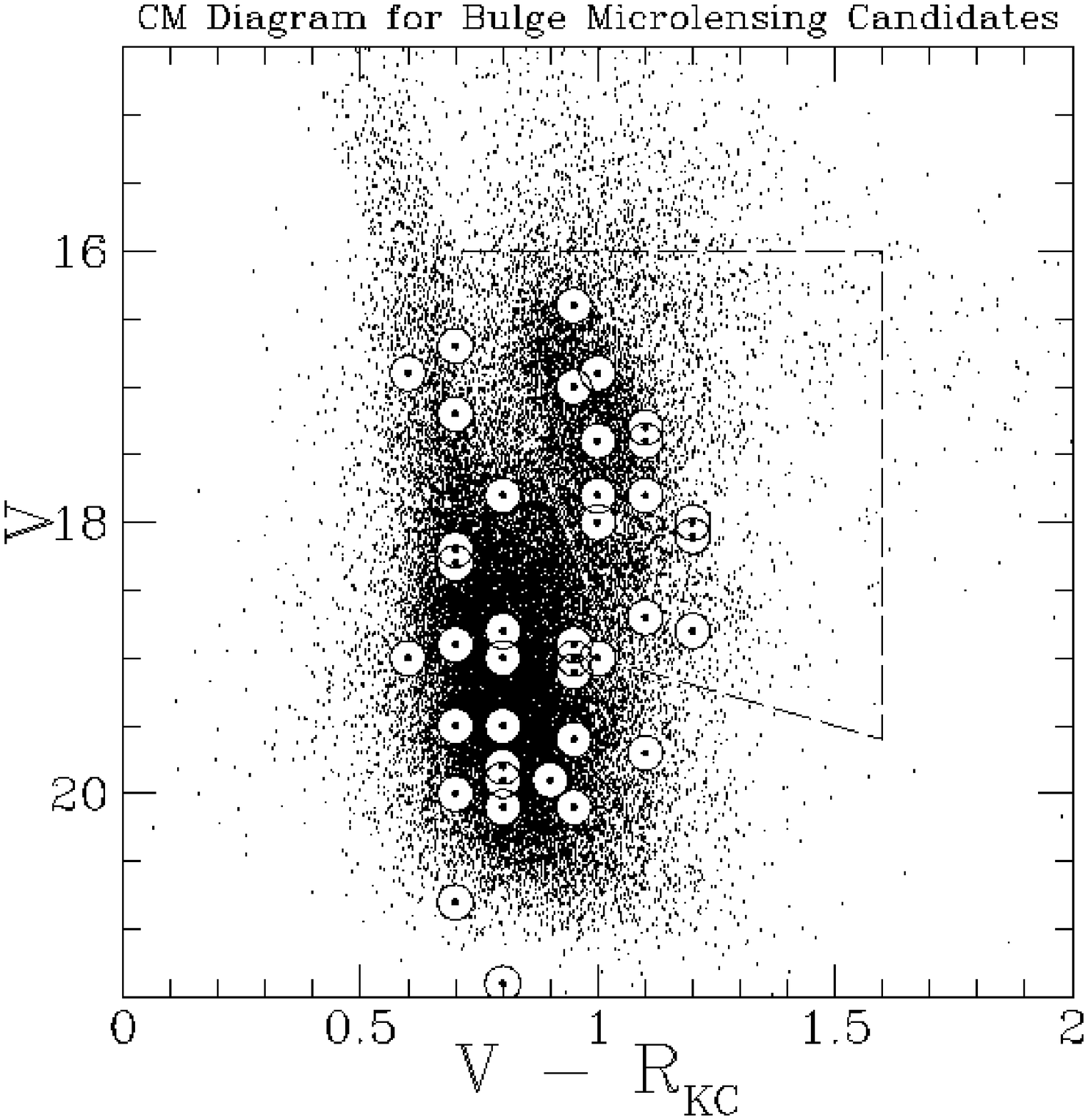}
\end{minipage}
\caption{
The left panel shows a color-magnitude diagram (CMD) made from
portions of 9 MACHO LMC fields with the position of the source
stars for the 8 microlensing candidate events from our first
two years marked.
The right panel shows a CMD created from portions of
24 fields in the bulge showing the source star positions for the 42
microlensing candidates from 1993.  
}\label{microcmd}
\end{center}
\end{figure}

We have recently concluded an analysis of the first two years of data on
22 fields in the LMC.  These data have yielded nine  microlensing candidates
eight of which are shown in Figure~\ref{lmcmicro}.  
The position of the baseline 
color and magnitude for these 8 events is shown
in a composite color-magnitude diagram of LMC data in Figure~\ref{microcmd}.
The ninth event occurred in an unresolved source star which was lensed during
the template observation.
Because our efficiency analysis does account for sources which are not
represented by objects in our template, this event is not considered further.
The measured optical depths for various samples in both
the LMC and the bulge are shown in Table 1.

A fraction more than one microlensing event would be 
expected from known populations in the
Milky Way and the LMC after monitoring 8.6 x $10^6$ stars for two years.
Thus, we considered two samples for analysis, a six event sample
(excluding one possible LMC-LMC event and one additional event) and an
eight event sample.  Both samples produce similar conclusions.  We 
detect about half the predicted optical depth if the dark halo were
composed of only Machos.
A maximum likelihood analysis suggests that we are detecting $\sim${50\%} 
of the dark halo mass in objects of $\sim$~{0.5 M$_{\odot}$}. 
A full account of this analysis is forthcoming \cite{alc96}.
This analysis and a companion analysis 
looking for evidence of short
microlensing events in our data \cite{spike} support and extend the
EROS negative result \cite{erosshort1} demonstrating the 
absence of short microlensing events.  Our result limits objects in
the mass range $2.5 \times 10^{-6} - 8.0 \times  10^{-2} M_{\odot}$
(Venus through brown dwarf masses)
to contributing less than 20\% of the dark halo.

Although the MACHO Project has only analyzed one season of bulge data 
to date, this season yielded 42 microlensing candidates.  We have 
detected an additional 75 bulge events with our real time  alert system 
since late 1994.
We have used both the full 1993
sample and a subset derived from bright, clump giant source stars to
calculate an optical depth to the bulge.  Because detailed efficiency
simulations have not been completed for our bulge data, the corrections
for faint source stars are somewhat uncertain, but the clump giant sample
is not subject to significant efficiency corrections.  Both samples
yield a higher optical depth than the expected $1 \times 10^{-6}$ optical
depth to the bulge \cite{kim91} \cite{pac91}. OGLE and DUO
report similar measurements of optical depth \cite{ogle} \cite{duo}.
These data support the hypothesis of a massive bar
in the Milky Way and may also provide evidence
of significant disk dark matter
\cite{oglebar} \cite{bulge45}.  Microlensing of apparent disk stars as
evidenced by the microlensing events along the upper blue sequence in
Figure~\ref{microcmd} underscores the possibility of significant disk
dark matter.  

\begin{figure}[h!]
\begin{center}
\epsfxsize=.45\columnwidth
\begin{minipage}{\epsfxsize}
\epsfbox{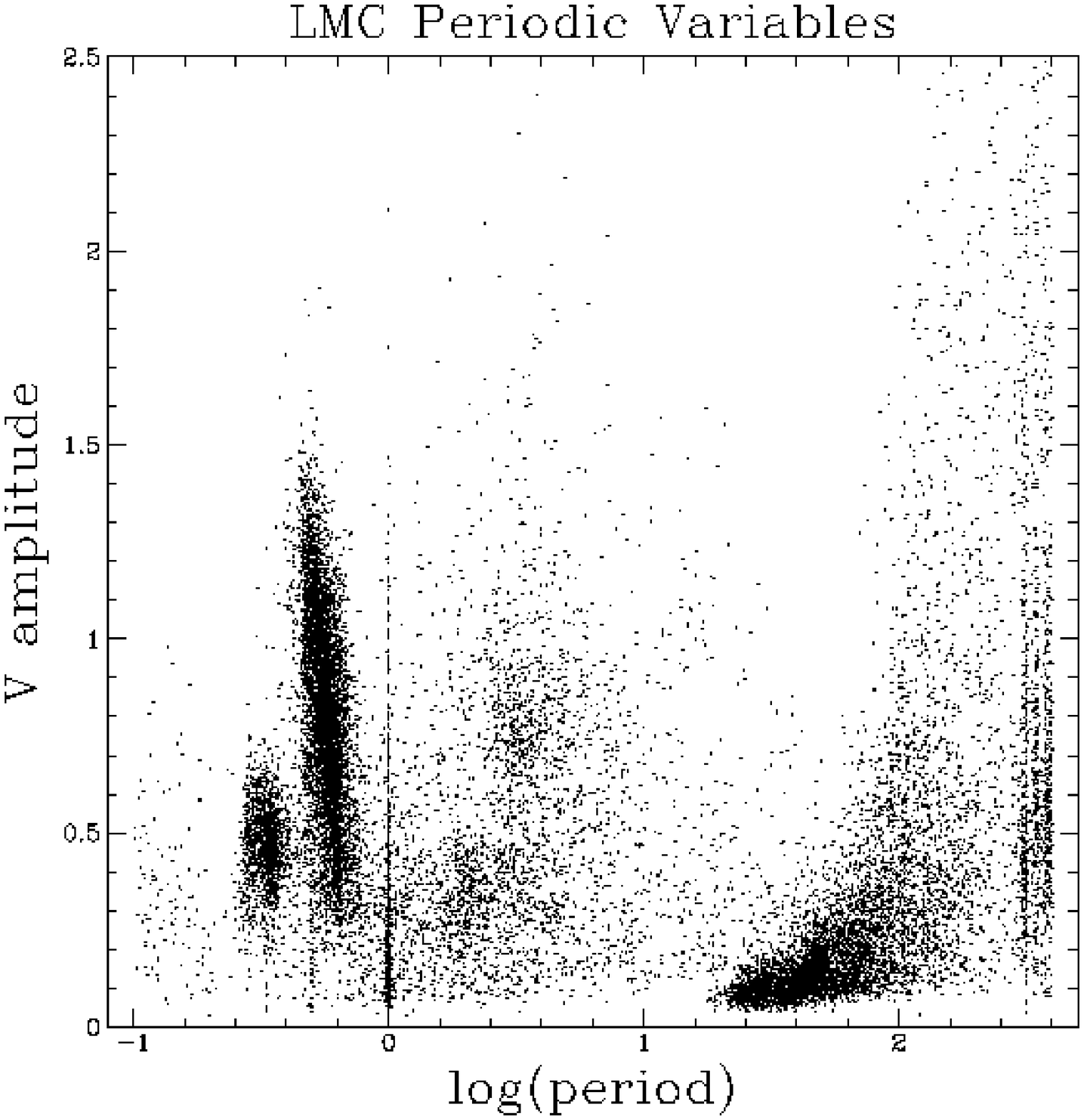}
\end{minipage}
\begin{minipage}{\epsfxsize}
\epsfbox{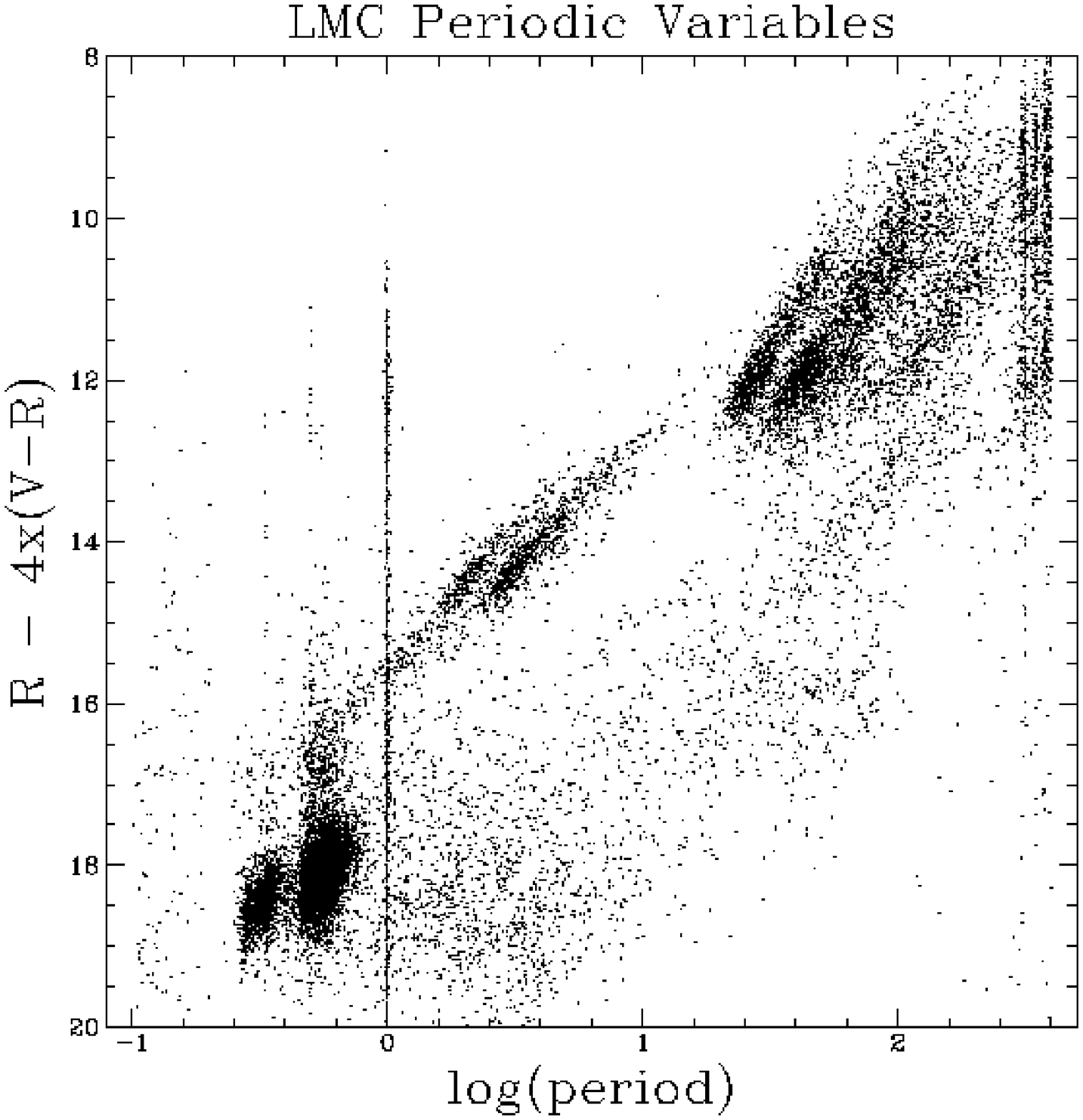}
\end{minipage}
\caption{
These panels plot the period versus the amplitude (left) and period
versus a reddening corrected magnitude (right) for the 23,589 periodic
variables found in the central 11 square degrees of the LMC.
}\label{lmcvars}
\end{center}
\end{figure}

\section{Variables in the MACHO Database}

\begin{figure}
\begin{center}
\epsfxsize=.45\columnwidth
\begin{minipage}{\epsfxsize}
\epsfbox{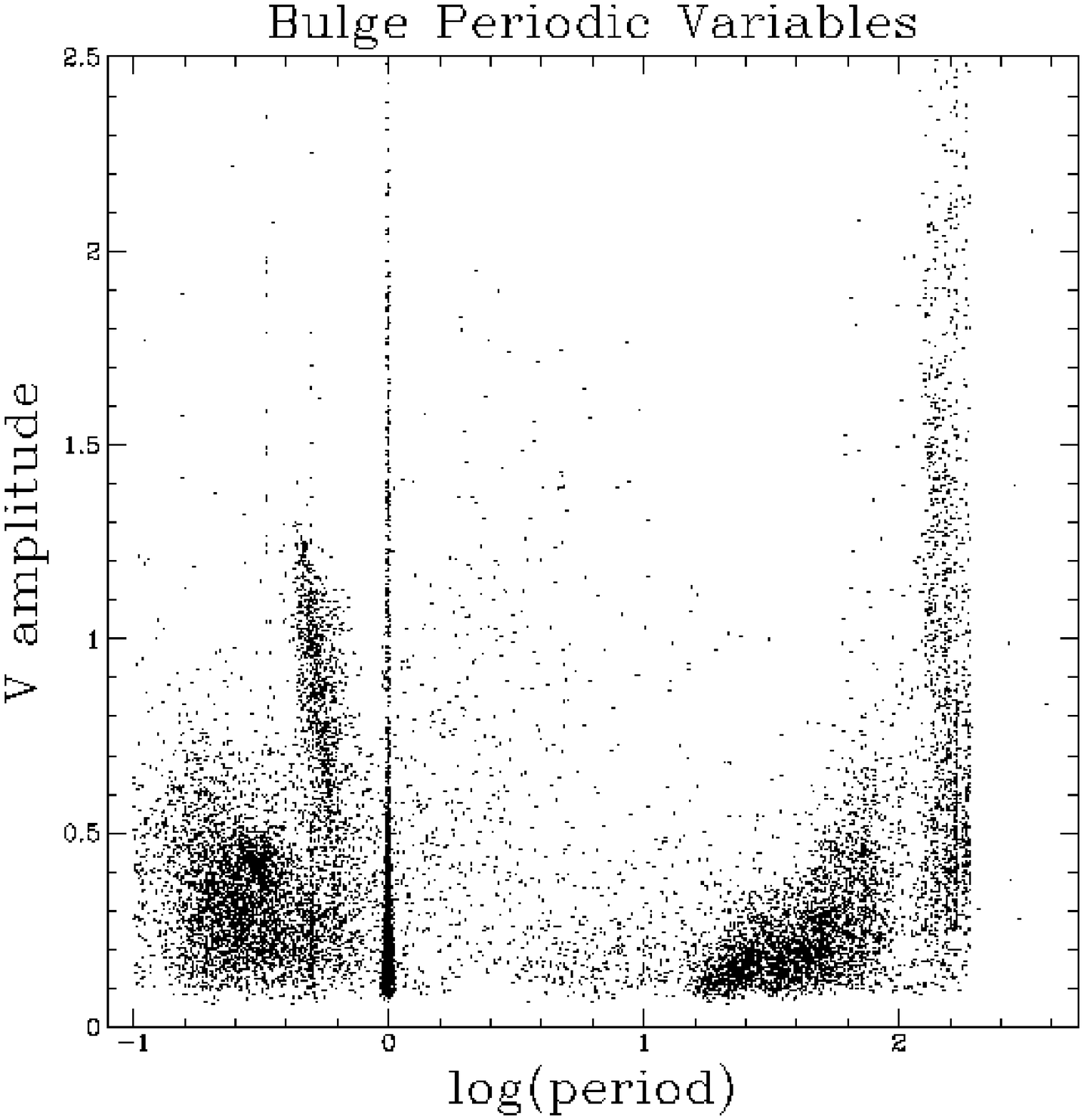}
\end{minipage}
\begin{minipage}{\epsfxsize}
\epsfbox{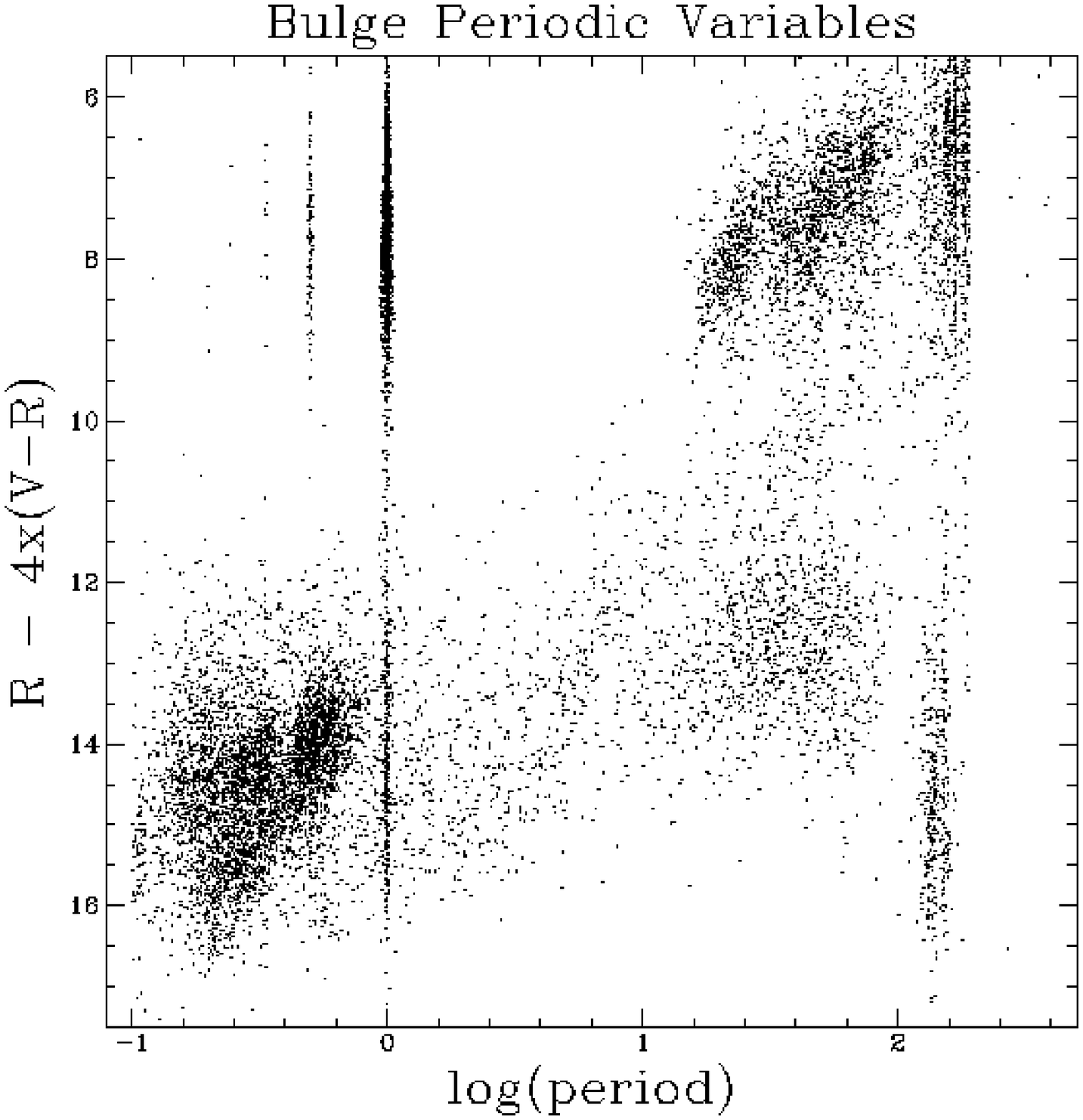}
\end{minipage}
\caption{
These panels plot the period versus the amplitude (left) and period
versus a reddening corrected magnitude (right) for the 14,776 periodic
variables found in the 12 square degrees of the bulge monitored in 1993.
}\label{blgvars}
\end{center}
\end{figure}

\begin{figure}
\begin{center}
\epsfxsize=.3\columnwidth
\begin{minipage}{\epsfxsize}
\epsfbox{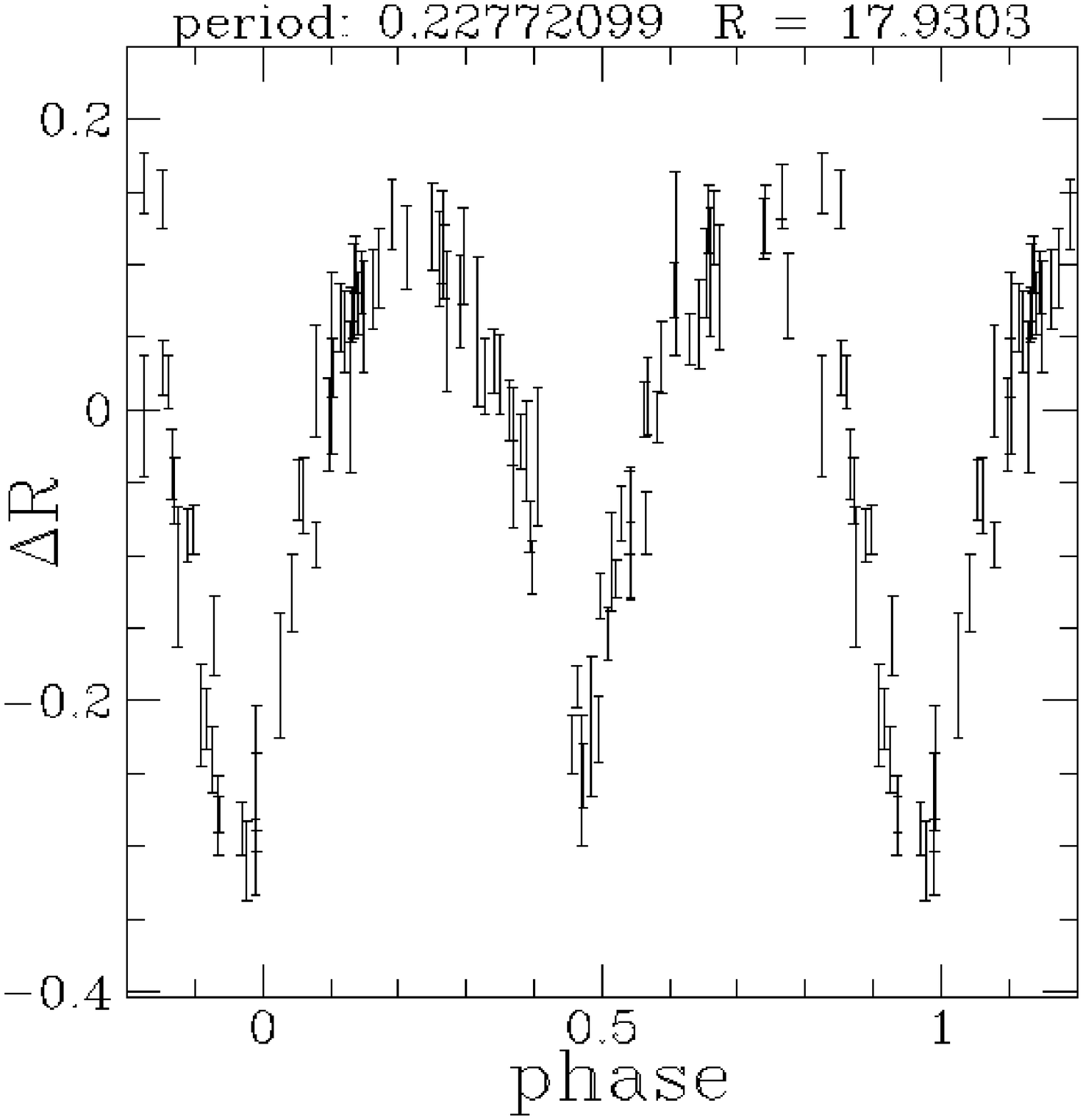}
\end{minipage}
\begin{minipage}{\epsfxsize}
\epsfbox{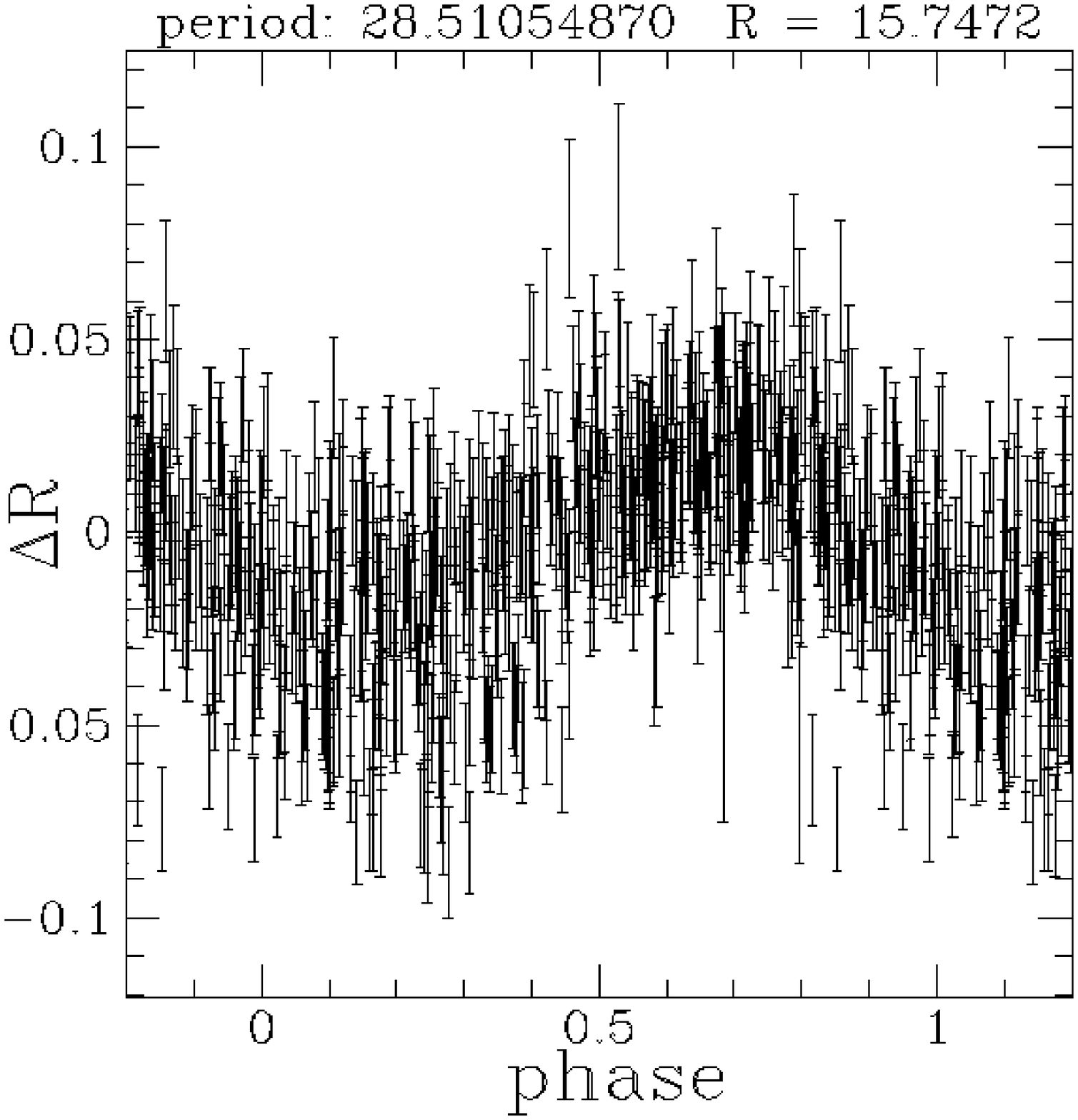}
\end{minipage}
\begin{minipage}{\epsfxsize}
\epsfbox{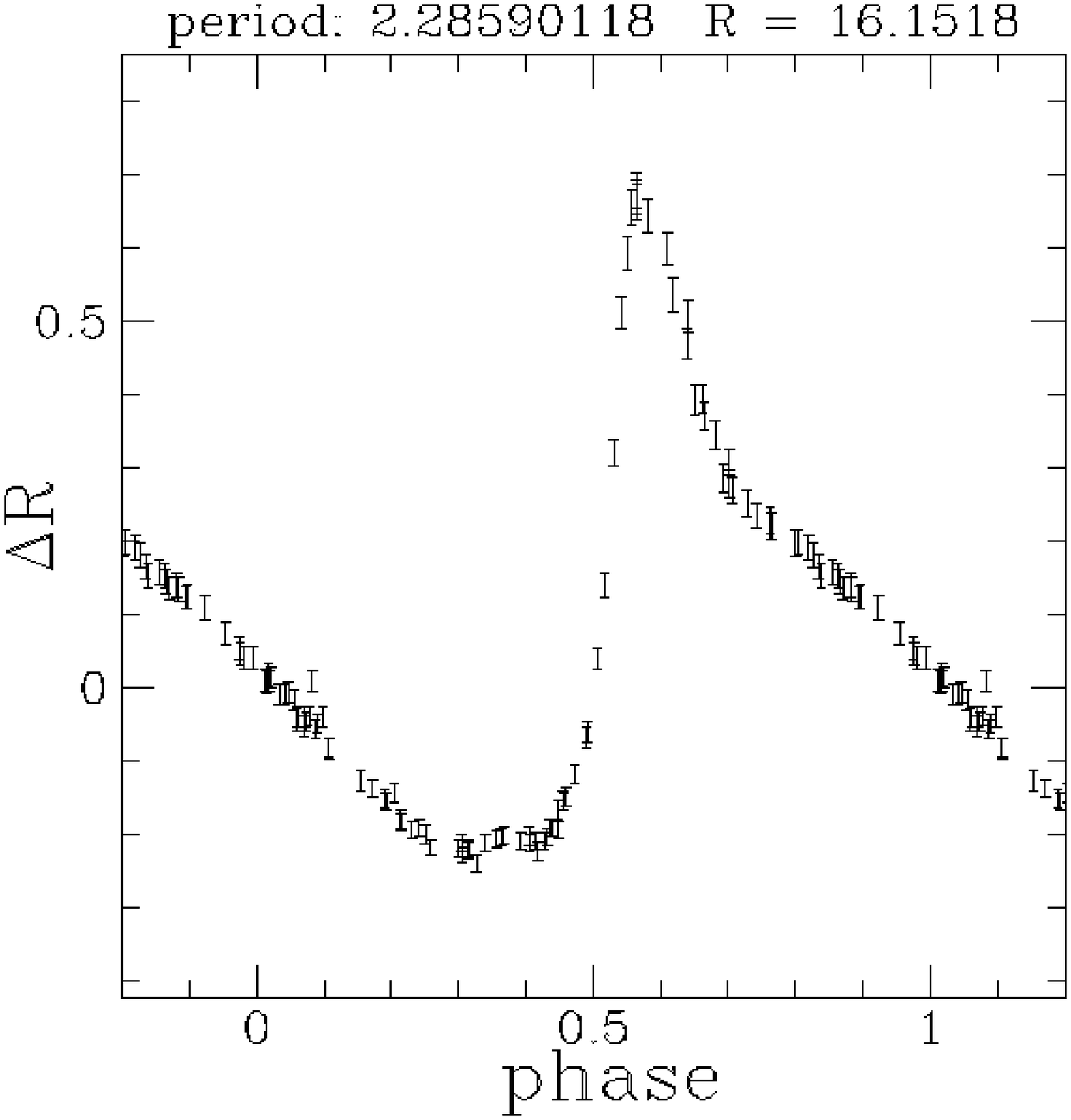}
\end{minipage}
\begin{minipage}{\epsfxsize}
\epsfbox{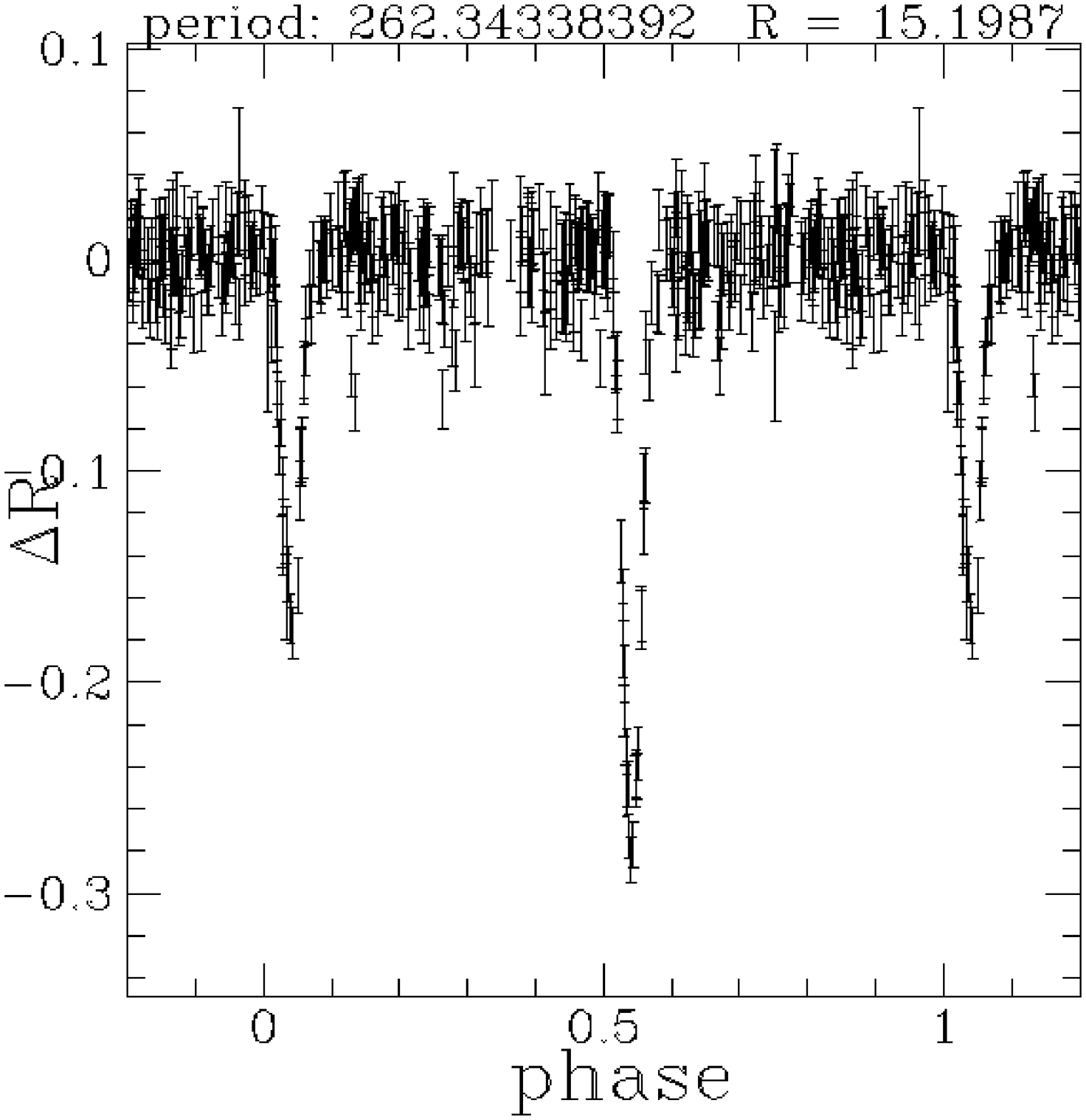}
\end{minipage}
\begin{minipage}{\epsfxsize}
\epsfbox{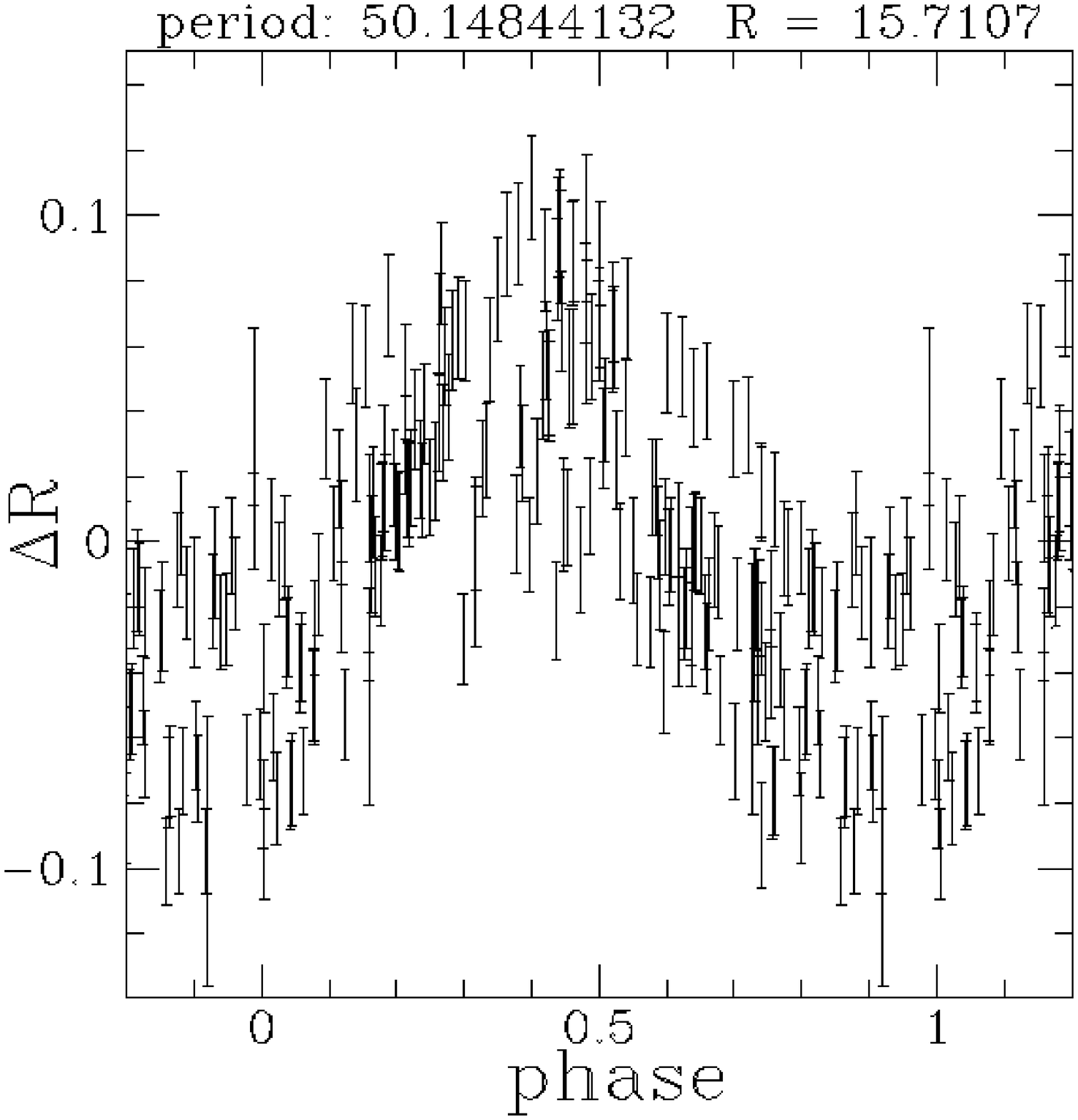}
\end{minipage}
\begin{minipage}{\epsfxsize}
\epsfbox{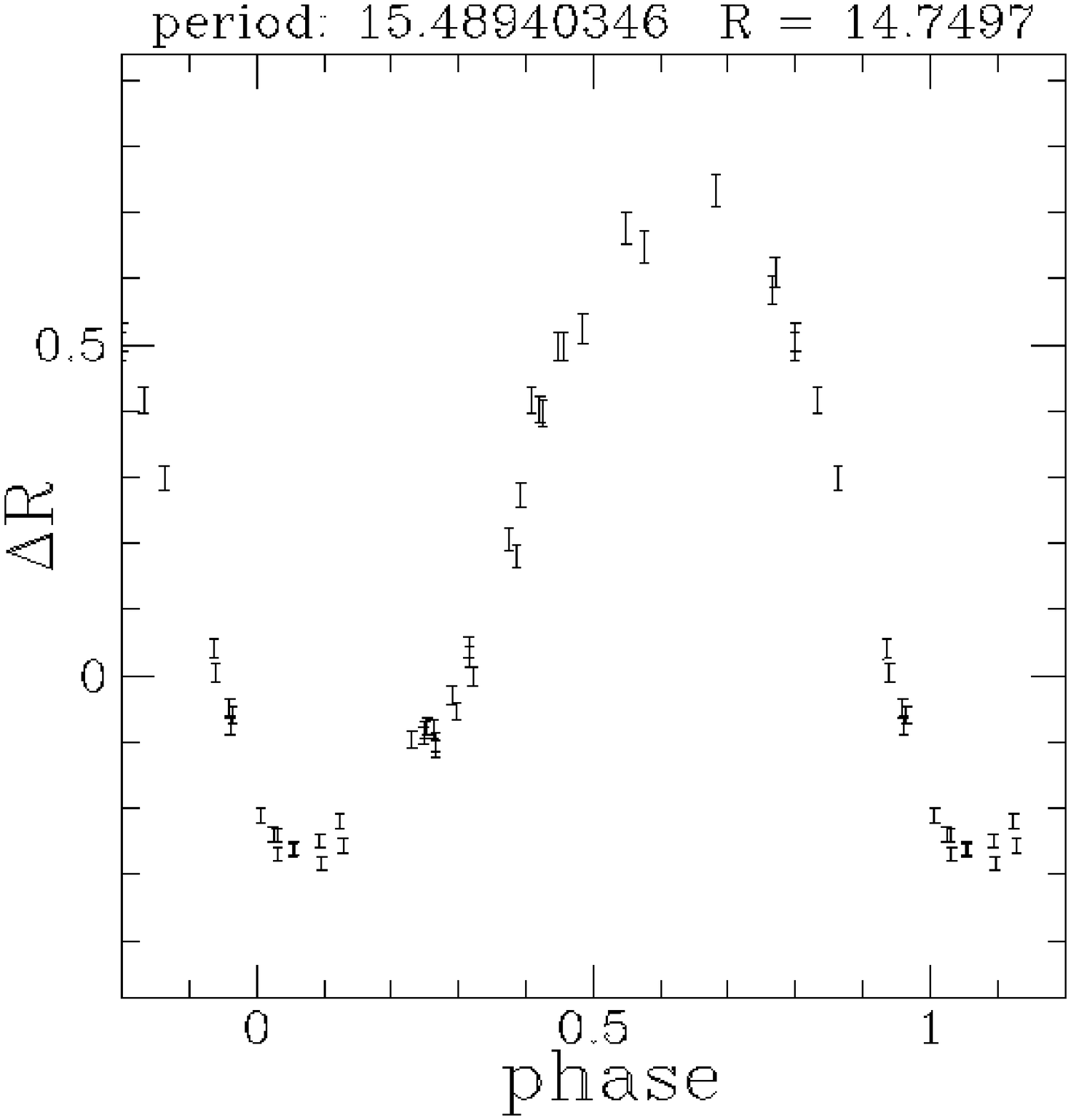}
\end{minipage}
\caption{
These panels show phased light curves of periodic variables in the MACHO
database.  The mean magnitude and period is listed at the top of the panel
and the plot is of phase versus change in R magnitude.
The left panels show a W Uma toward the bulge (top) and an
eclipsing red giant pair in the LMC bar (bottom).  The middle panels show
examples from the two red giant period-luminosity tracks.  The right panels
show a BL Her and a W Vir in the bulge.
}\label{varfig}
\end{center}
\end{figure}

The color-magnitude diagrams (CMDs)of Figure~\ref{microcmd} demonstrate the
photometry product of the microlensing search.  The mean photometric error
at V = 20 for hundreds of epochs is about 25{\%} for the LMC and about 
40{\%} for the bulge.  The LMC CMD is created from less than 1{\%} of
the LMC light curves while the bulge CMD is created from about 1{\%} of the
bulge light curves.
Much of the scatter in these diagrams is due to
the uncertainty in the transformations as discussed earlier.
As part of the analysis of the light curves in the search for microlensing,
a variability statistic is calculated for each light curve.  
The $\chi^2$/dof for a fit to a
constant is calculated using only the 80\%  of points closest
to the median for each light curve.  The light curve is defined as variable
if this, robust, $\chi^2$/dof is greater than the regular $\chi^2$/dof which
can occur by chance 1\%  of the time for a constant light curve.  The nightly
sampling for extended periods of time results in an effective sampling
even for variables with periods much shorter than a day.  This is because
the sampling interval constantly changes since we attempt to observe
each field at its lowest airmass in a night.

Of the $\sim$ 9 million LMC light curves from
the first two years of LMC data 37,051 light curves
were identified as variable.  
Of the $\sim$ 12 million bulge light curves
currently analyzed from 1993, 38,718 were identified as variable.  
An additional analysis of LMC
light curves was conducted to identify those which were rejected because
of relatively rare excursions below the robust envelope defined above.
This analysis identified about 1000 more eclipsing binaries, 3000 more
low amplitude RR Lyrae (primarily due to confusion), and about 6000
low amplitude, low signal-to-noise variable candidates. These 85,000
light curves were tested for periodic behavior using a code designed
specifically for the MACHO Project \cite{reimann94}.  When the red and
blue light curve yielded the same period to within 0.1 \% the
light curve was defined as representing a periodic variable.  
Figure~\ref{lmcvars} shows period-amplitude and period-magnitude
plots for the 23,589 periodic variables identified in the LMC.
Figure~\ref{blgvars} shows the period-amplitude and period-magnitude
plots for the 14,776 periodic variables identified in the bulge.
A preliminary description of the LMC variables has been presented
\cite{cook}.

These figures hold a wealth of information about the variable 
stellar populations
of the LMC and the bulge.  There are also certain systematic features 
in this data set which are apparent.  There are vertical lines in the
distributions for two reasons due to the phasing code we have used.  
The phasing code will return an alias of an integer fraction of a day 
when it tries to phase a low signal-to-noise (low amplitude) aperiodic
variable.  For high signal-to-noise aperiodic or very long period
variables, it will fold them with a
period of the light curve length.  There are multiple, long period vertical
lines because different fields yielded different light curve durations.
We have used these figures to define regions in period,
magnitude, amplitude space which contain highly purified samples of
particular types of variables.  We have also used the blue-to-red amplitude
ratio to discriminate pulsating from eclipsing variables \cite{dante96}.
One can clearly see the RRab and RRc in
the LMC as distinct clumps near log(period) = -0.3 and -0.5, respectively.
In the bulge, the RRc locus is clearly contaminated by short period, low
amplitude variables, mostly eclipsing binaries.  The fundamental and
overtone Cepheid period-luminosity (PL) loci 
stand out in the LMC data and are absent in the
bulge data as would be expected because of its mostly old population.  Perhaps
the most striking feature of these diagrams is the appearance of clear,
multiple PL relations for bright, long period variables.  These are all
giants lying just beyond the tip of the giant branch.  As can be seen,
they have low amplitudes with a suggestion of increasing amplitude with
increasing period.  Eclipsing binaries are found scattered across these
diagrams.
More detailed discussions of 
the $\sim$ 10,000 RR Lyrae \cite{dante96}, 1500
Cepheids in the LMC \cite{doug96}, and Type II Cepheids and RV Tauri
in the LMC \cite{pollard96} contained in the MACHO database are presented
elsewhere in this volume.  

Examples of some of the more interesting periodic variables from our 
database are shown in Figure~\ref{varfig}.  Eclipsing binaries from
the EROS and OGLE survey are discussed in detail elsewhere in this 
volume \cite{ruc96} \cite{tobin96} \cite{pac96}.  
Examples of extremes in period for eclipsing systems in the 
MACHO data are shown in the left panels.  About a dozen eclipsing,
red giant pairs with periods over 100 days have been identified in 
our LMC data so far.  Two examples of the low amplitude, periodic
red giant variables with periods less than 100 days are shown in the middle
panels.  There seems to be quite a range in the constancy of the
amplitude of stars in this region of the period-amplitude diagram.
The relatively short periods and the sinusoidal light
curves are suggestive of overtone oscillations.  Although there are
few classical Cepheids along our line of sight to the bulge, we have
discovered about 20 new BL Her and W Vir stars.  Examples of two of these
are shown in right panels. 

\begin{figure}
\begin{center}
\epsfxsize=.3\columnwidth
\begin{minipage}{\epsfxsize}
\epsfbox{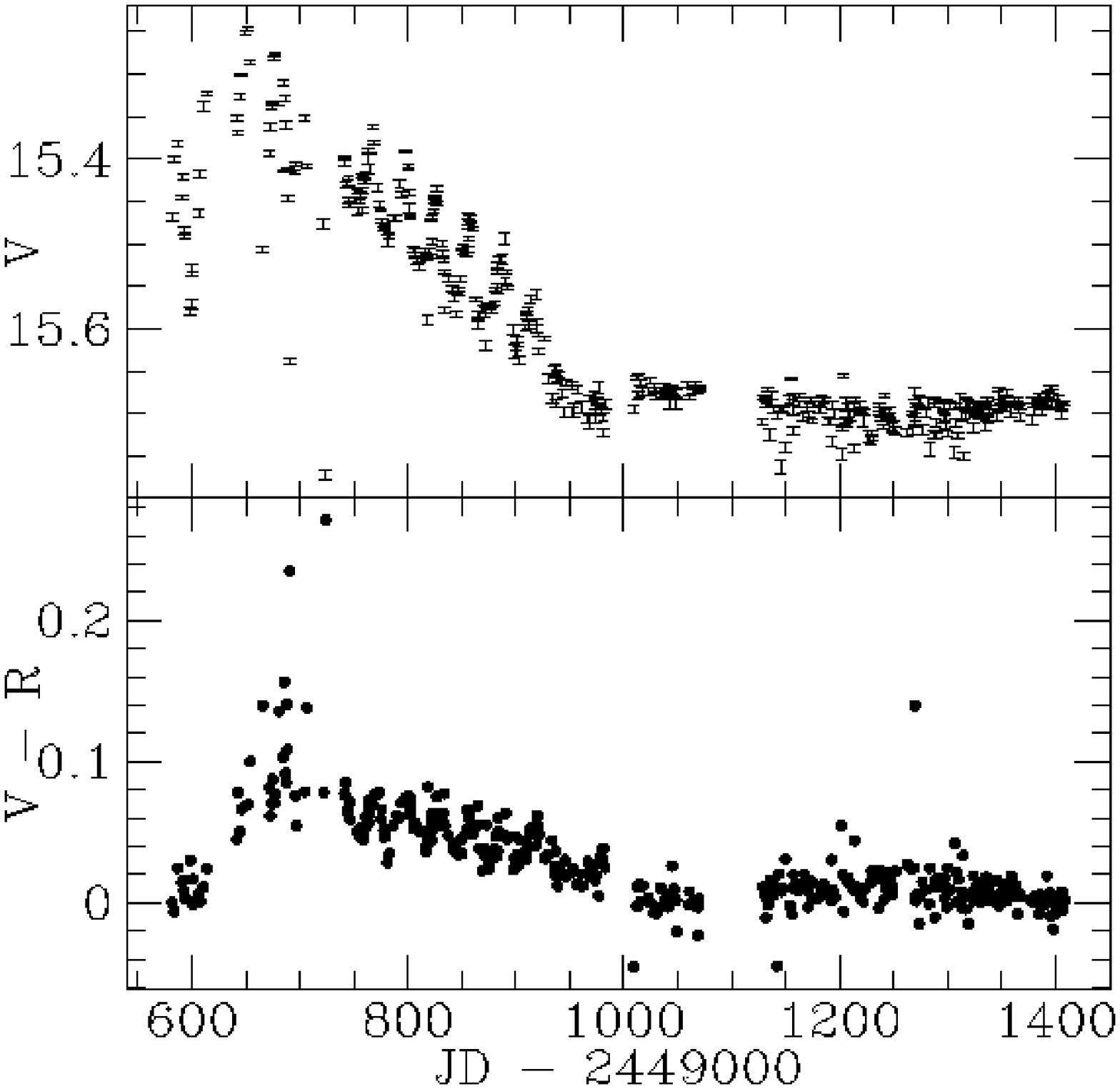}
\end{minipage}
\begin{minipage}{\epsfxsize}
\epsfbox{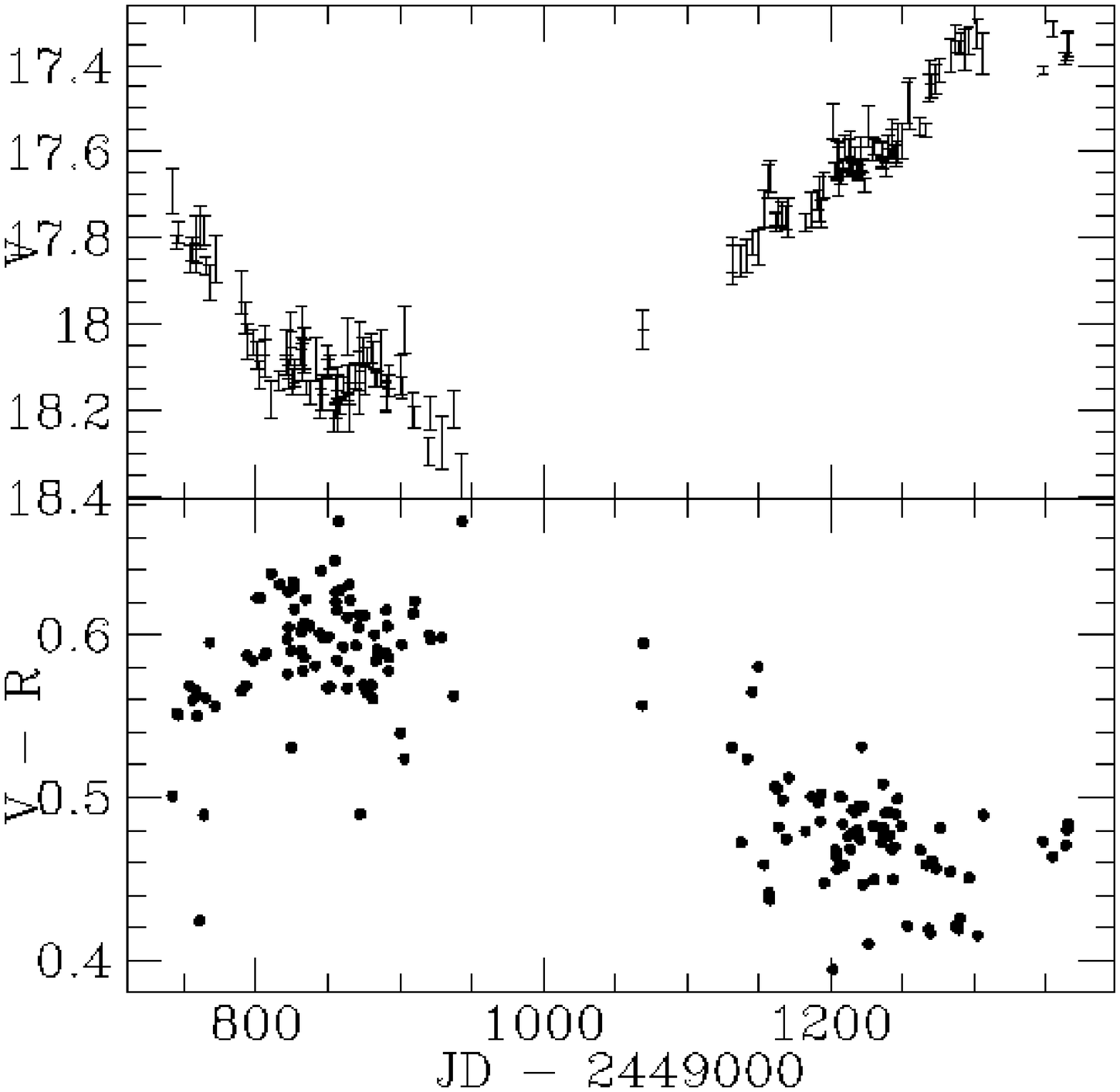}
\end{minipage}
\begin{minipage}{\epsfxsize}
\epsfbox{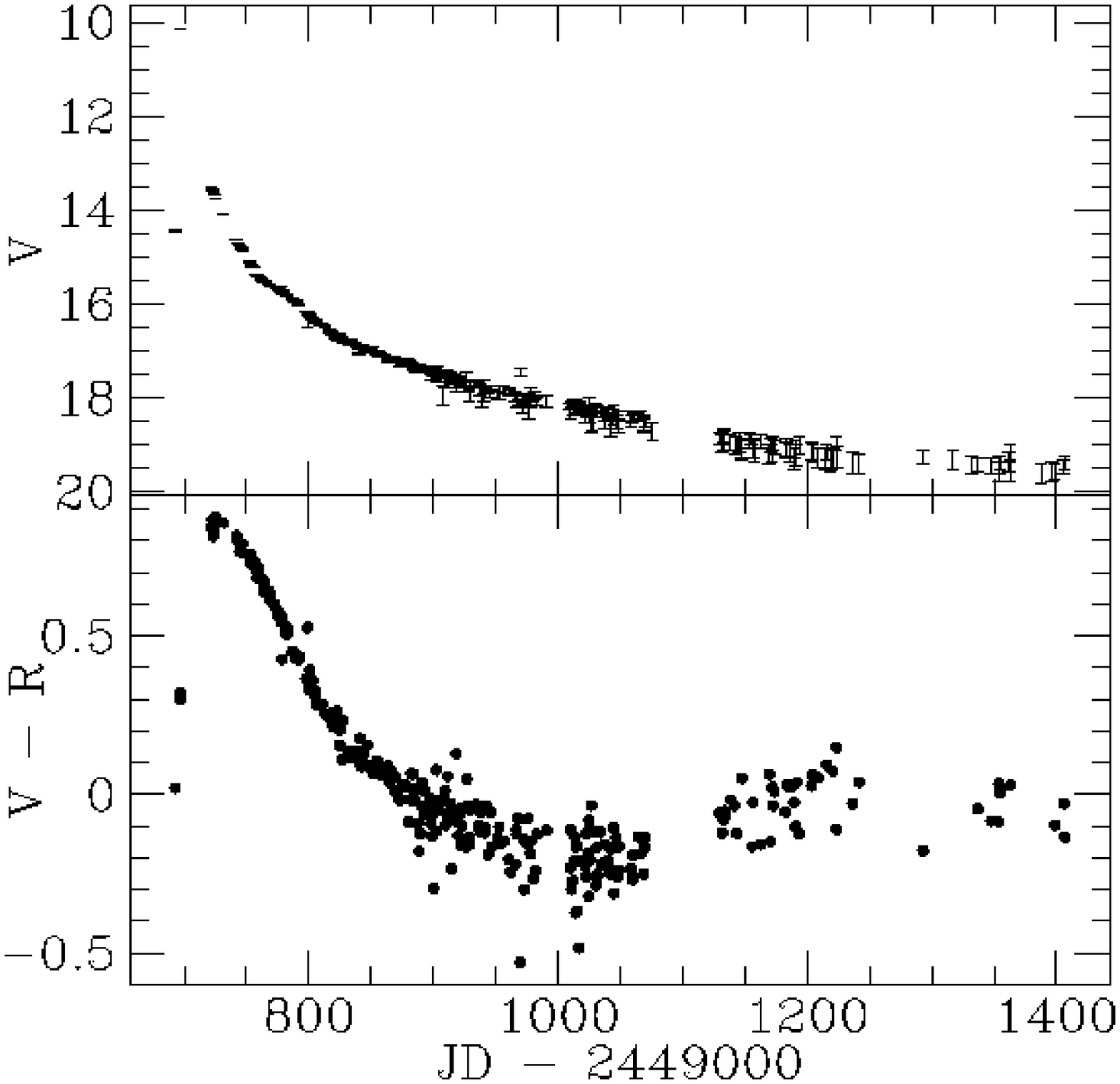}
\end{minipage}
\caption{
These are interesting light curves from the LMC aperiodic blue variable
catalog.  The V light curve is plotted over the ($V - R$) light curve.
The left panel shows a star oscillating less as it declines in brightness
until it is roughly constant.  The middle panel shows the light curve of
a z = 0.15 Seyfert behind the bar of the LMC \protect\cite{seyfert}.  The right panel shows
the light curve of a nova in the LMC in November of 1992 \protect\cite{nova92}.
}\label{bump}
\end{center}
\end{figure}

We have also begun a detailed examination of the aperiodic blue variables
found in the LMC data.  Our interest in these stars was stimulated by
the observation of rare, small amplitude outbursts in mostly constant
stars which we have termed bumpers.  We have compiled a catalog of
about 400 aperiodic blue variables which demonstrate some truly
odd photometric behavior.  These stars were chosen from about 250,000
light curves with $14.0 < V < 18.0$ and $(V-R) < 0.3$ \cite{purdue}.
Characteristic light curves of two major classes of variability found in  
the catalog have been previously presented \cite{cook}.
Figure~\ref{bump}
shows three other interesting blue, aperiodic variables: an odd one,
a background Seyfert and a nova. 
In this set of aperiodic blue variables we have also found 
supersoft X-ray sources, and planetary nebulae.

\section{Summary}

The unevenly spaced sampling, the wide field of view, the simultaneous
color information and the long, densely sampled light curves make the
MACHO database a treasure trove for the study of 
time variability of astrophysical
sources.  While we have only begun to examine this database outside the
context of the search for Machos, the early returns promise much for the
future.  The MACHO Project is also moving rapidly toward achieving 
its goal of determining the baryonic content
of the halo.  The MACHO Project plans to continue operation at
least through 1999.  At the end of this time, we expect 
to have a sample of about 50 microlensing events across the face of the LMC,
about 1000 events toward the bulge, and a database of light curves for
more than 50 million stars in the LMC, the SMC and the bulge, spanning 
8 years.

\acknowledgements{
We are very grateful for the skilled support by the technical staff at MSO.  
Work at LLNL is supported by DOE contract W7405-ENG-48. 
Work at the CfPA is supported NSF AST-8809616 and AST-9120005.  
Work at MSSSO is supported by the Australian Department of Industry, 
Technology and Regional Development.  
WJS is supported by a PPARC Advanced Fellowship.  
KG thanks support from DOE OJI, Sloan, and Cottrell awards.
CWS thanks support from the Sloan, Packard and Seaver Foundations.
}

\vfill
\end{document}